\documentclass{article}
\usepackage{spconf,amsmath,graphicx}
\usepackage{caption, booktabs} 
\usepackage{hyperref}
\title{LEARNING SINGING FROM SPEECH}
\name{Liqiang Zhang\sthanks{Work performed while interning at Tencent AI Lab}\textsuperscript{1}, Chengzhu Yu\textsuperscript{2}, Heng Lu\textsuperscript{2}, Chao Weng\textsuperscript{2}, Yusong Wu\textsuperscript{2},
Xiang Xie\textsuperscript{1}, Zijin Li\textsuperscript{3}, Dong Yu\textsuperscript{2}}

\address{\textsuperscript{1}Beijing Institute of Technology \quad \textsuperscript{2}Tencent AI Lab   \quad \textsuperscript{3}China Conservatory of Music\\
\texttt{\small{\{zhlq,xiexiang\}@bit.edu.cn,\{czyu,bearlu,cweng,ysw,dyu\}@tencent.com,lizijin2019@hotmail.com}}}

\begin{document}

\maketitle
\begin{abstract}
We propose an algorithm that is capable of synthesizing high quality target speaker's singing voice given only their normal speech samples. The proposed algorithm first integrate speech and singing synthesis into a unified framework, and learns universal speaker embeddings that are shareable between speech and singing synthesis tasks. Specifically, the speaker embeddings learned from normal speech via the speech synthesis objective are shared with those learned from singing samples via the singing synthesis objective in the unified training framework. This makes the learned speaker embedding a transferable representation for both speaking and singing. We evaluate the proposed algorithm on singing voice conversion task where the content of original singing is covered with the timbre of another speaker's voice learned purely from their normal speech samples. Our experiments indicate that the proposed algorithm generates high-quality singing voices that sound highly similar to target speaker’s voice given only his or her normal speech samples. We believe that proposed algorithm will open up new opportunities for singing synthesis and conversion for broader users and applications.
\end{abstract}
\begin{keywords}
Singing Synthesis, Singing Voice Conversion
\end{keywords}
\section{Introduction}
\label{sec:intro}

Singing is one of the most important music expression and the techniques of singing synthesis have many applications in entertainment industries. Over the past decades, many approaches have been proposed for singing synthesis. These include methods based on concatenative unit selection \cite{bonada2016expressive} as well as more recent approaches based on deep neural network (DNN) \cite{nishimura2016singing} and autoregressive generation models \cite{blaauw2017neural}. 

While existing singing synthesis algorithms are capable of producing natural singing, it normally requires a large amount of singing data for training new voices. Compared to normal speech data, singing data is much more difficult and expensive to collect. To address such limitation, more data efficient singing synthesis approaches \cite{blaauw2019data} have been proposed recently, which adapts a multi-speaker trained singing synthesis model with a small amount of target speaker's singing data. 

Alternatively, singing synthesis with new voices can be achieved through singing voice conversion. The task of singing voice conversion is to convert one’s singing with the voice of another while keeping singing content the same. Traditional singing voice conversion \cite{kobayashi2014statistical,kobayashi2015statistical,villavicencio2010applying} relies on parallel singing data to learn conversion function between different speakers. However, a recent study \cite{nachmani2019unsupervised} on unsupervised singing voice conversion uses a WaveNet \cite{oord2016wavenet} based autoencoder architecture to achieve singing voice conversion without parallel singing data or even the transcribed lyrics or notes. 

While data efficient singing synthesis approach \cite{blaauw2019data} as well as unsupervised singing voice conversion method \cite{nachmani2019unsupervised}, could efficiently generate singing with new voices, it still requires a minimal amount of singing voice samples from target speakers. This has limited the applications of singing voice synthesis to relatively restricted scenarios where the target speaker's singing voice has to be available. 

On the other hand, normal speech samples are much easier to collect than singing. However, there are only a few studies have investigated the use of speech samples for singing synthesis. The speech-to-singing synthesis method proposed in \cite{saitou2007speech} attempts to convert a speaking voice to singing by directly modifying acoustic features such as f0 contour and phoneme duration in read speech. While speech-to-singing approaches could produce singing from read lyrics, it normally requires non-trivial amount of manual tuning of acoustic features for achieving high intelligibility and naturalness of singing voices. 

\begin{figure*}[t]
  \centering
  \includegraphics[clip, trim=0.6cm 0.8cm 0.6cm 0.8cm, width=0.9\textwidth]{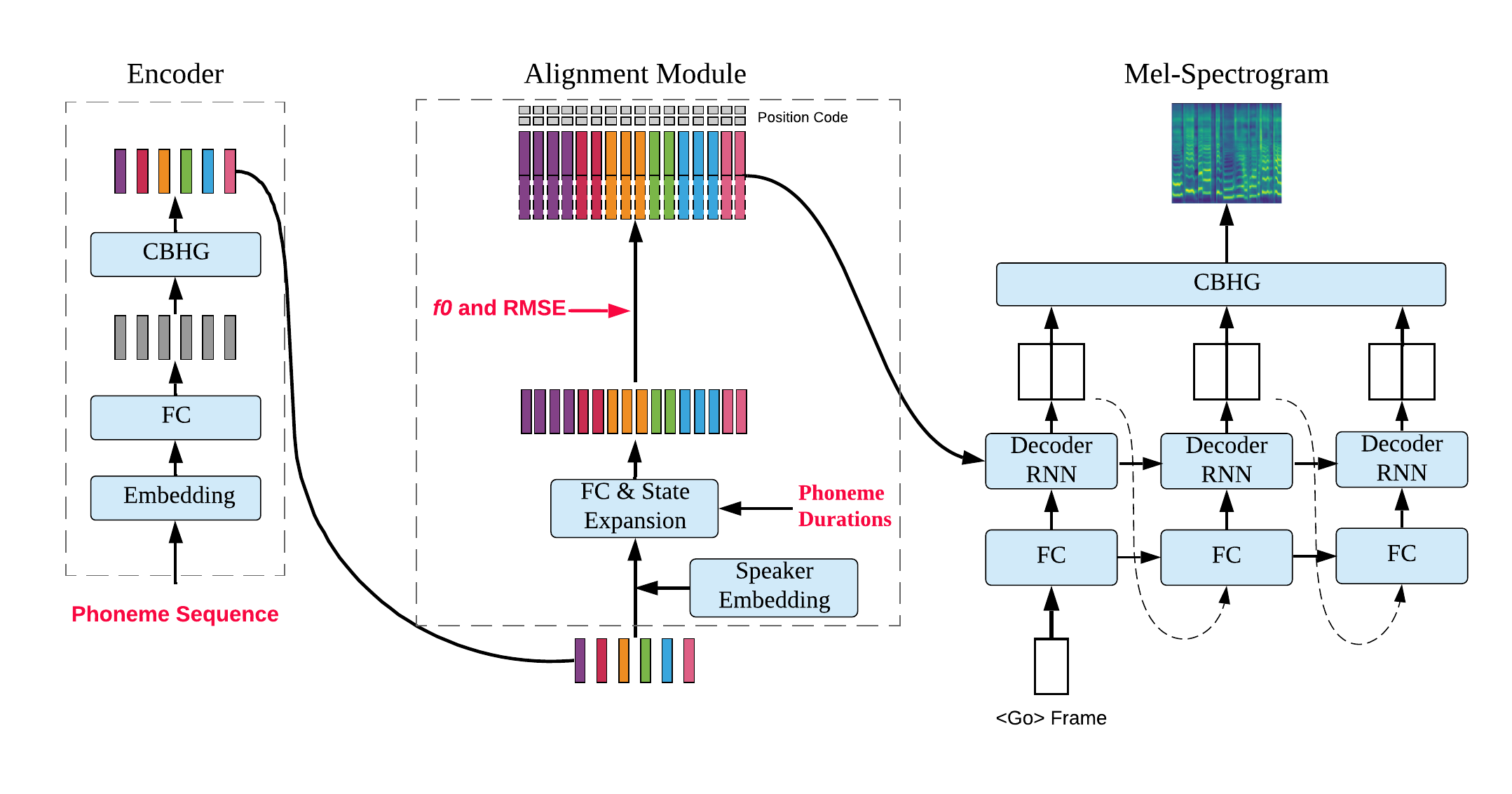}
   \caption{Model architecture of DurIAN-4S.}
\vspace{-3mm}
\label{fig:f2}
\end{figure*}

In this paper, we propose an algorithm that directly synthesizes natural singing with target speakers' voice by learning their voice characteristics from speech samples\footnote{Sound demo of proposed algorithm can be found at \url{https://tencent-ailab.github.io/learning_singing_from_speech}}. The key part of proposed algorithm is to learn universal speaker embeddings, such that the speaker embeddings learned for the task of speech synthesis can be used for singing synthesis, and vice versa. For this purpose, we use our recently proposed autoregressive generation model, Duration Informed Attention Network (DurIAN) \cite{yu2019durian}, for unifying text-to-speech and singing synthesis into a single framework. 
DurIAN, originally proposed for the task of multimodal synthesis, is essentially an autoregressive feature generation framework that could generates acoustic features (e.g., mel-spectrogram) from any audio source frame by frame. In proposed method, phoneme duration, fundamental frequency (F0) and root-mean-square energy (RMSE) are extracted from training data containing both singing or normal speech, and used as inputs for reconstructing target acoustic features. The entire model is trained jointly with learnable speaker embeddings as conditional input to the model. The trained model and speaker embeddings can be used to convert any singing into target speaker's voice by using his or her speaker embedding as conditional input.  

The paper is organized as following. Section 2 introduces the architecture of our conversion model. Section 3 introduces the experiment. Section 4 and 5 are the conclusion and acknowledgements.

\section{MODEL ARCHITECTURE}
\label{sec:pagestyle}
In this section, we first describe DurIAN based Speech and Singing Synthesis System (DurIAN-4S), a unified speech and singing synthesis system based on DurIAN. After that, we present singing voice conversion approach based on DurIAN-4S.

\subsection{DurIAN-4S}
\label{ssec:subhead}
While DurIAN was originally proposed for the task of multimodal speech synthesis, it is a general autoregressive framework that can be used for other synthesis tasks. The original DurIAN model is modified here to perform speech and singing synthesis at the same time. The major difference of DurIAN-4S compared to DurIAN is that it takes additional inputs. These additional inputs are attributes of singing that are useful for singing synthesis (music note, \textit{f0}, etc.). As the focus of this study is singing voice conversion\footnote{For the task of singing synthesis from note and lyrics, the note of music can be used as additional inputs.}, we use frame level \textit{f0} and root mean square energy (RMSE) extracted from original singing/speech as additional inputs\footnote{The \textit{f0} and RMSE will not be available at inference time of speech synthesis to be used as additional inputs. But, our objective is singing voice conversion, and the model will not be used for speech synthesis inference.} (Fig.~\ref{fig:f2}). 

The architecture of DurIAN-4S is illustrated in Fig.~\ref{fig:f2}. It includes (1) an encoder that encodes the context of each phoneme, (2) an alignment model that aligns the input phoneme sequence and to target acoustic frames, (3) an autoregressive decoder network that generates target mel-spectrogram features frame by frame. 

\subsubsection{Encoder}
We use phoneme sequence $\mathrm{x_{1:N}}$ directly as input for both speech and singing synthesis. The output of the encoder  $\mathrm{h_{1:N'}}$ is a sequence of hidden states containing the sequential representation of the input phonemes as
\begin{equation}
\mathrm{h_{1:N}} = \mathrm{encoder}(\mathrm{x_{1:N}}),
\end{equation}
where $\mathrm{N}$ is the length of input phoneme sequences\footnote{The state skipping structures in DurIAN \cite{yu2019durian} is not used here as it is not a necessary component for singing synthesis or conversion.}.

\subsubsection{Alignment model} 
The purpose of alignment model is to generate frame aligned hidden states that will be used as input for autoregressive generation. Here, the output hidden sequence from encoder $\mathrm{h_{1:N'}}$ is first concatenated with speaker embedding $\mathrm{m_{s}}$ and followed by a fully connect layer used for dimension reduction as
\begin{equation}\label{eq3}
\mathrm{h_{1:N}^{'}= fc(h_{1:N} \vee m_{s})}
\end{equation}
where $\vee$ indicates concatenation and $m_{s}$ indicates the embedding of speaker $s$. The output hidden states after dimension reduction layer will be expanded according to the duration of each phoneme as  
\begin{equation}
\mathrm{e_{1:T}} = \mathrm{state\_expand}(\mathrm{h_{1:N}^{'}}, \mathrm{d_{1:N}}), 
\end{equation}
where $T$ is the total number of input audio frames. The state expansion is simply the replication of hidden states according to the provided phoneme duration. The duration of each phoneme is obtained from force alignments performed on input phonemes and acoustic features. The frame aligned hidden states $\mathrm{e_{1:T}}$ is then concatenated with frame level \textit{f0}, RMSE, and relative  position  of  every  frame  inside  each  phone.
\begin{equation}\label{eq5}
\mathrm{e_{1:T}^{'}=e_{1:T} \vee f_{1:T} \vee r_{1:T} \vee p_{1:T}} 
\end{equation}
where $\mathrm{f_{1:T}}$ and $\mathrm{r_{1:T}}$ represents \textit{f0} and RMSE for each frame respectively. And $\mathrm{p_{1:T}}$ is the position code of each frame.

\begin{figure}[t]
  \centering
  \includegraphics[clip, trim=0.2cm 0.2cm 0.2cm 0.2cm, width=0.5\textwidth]{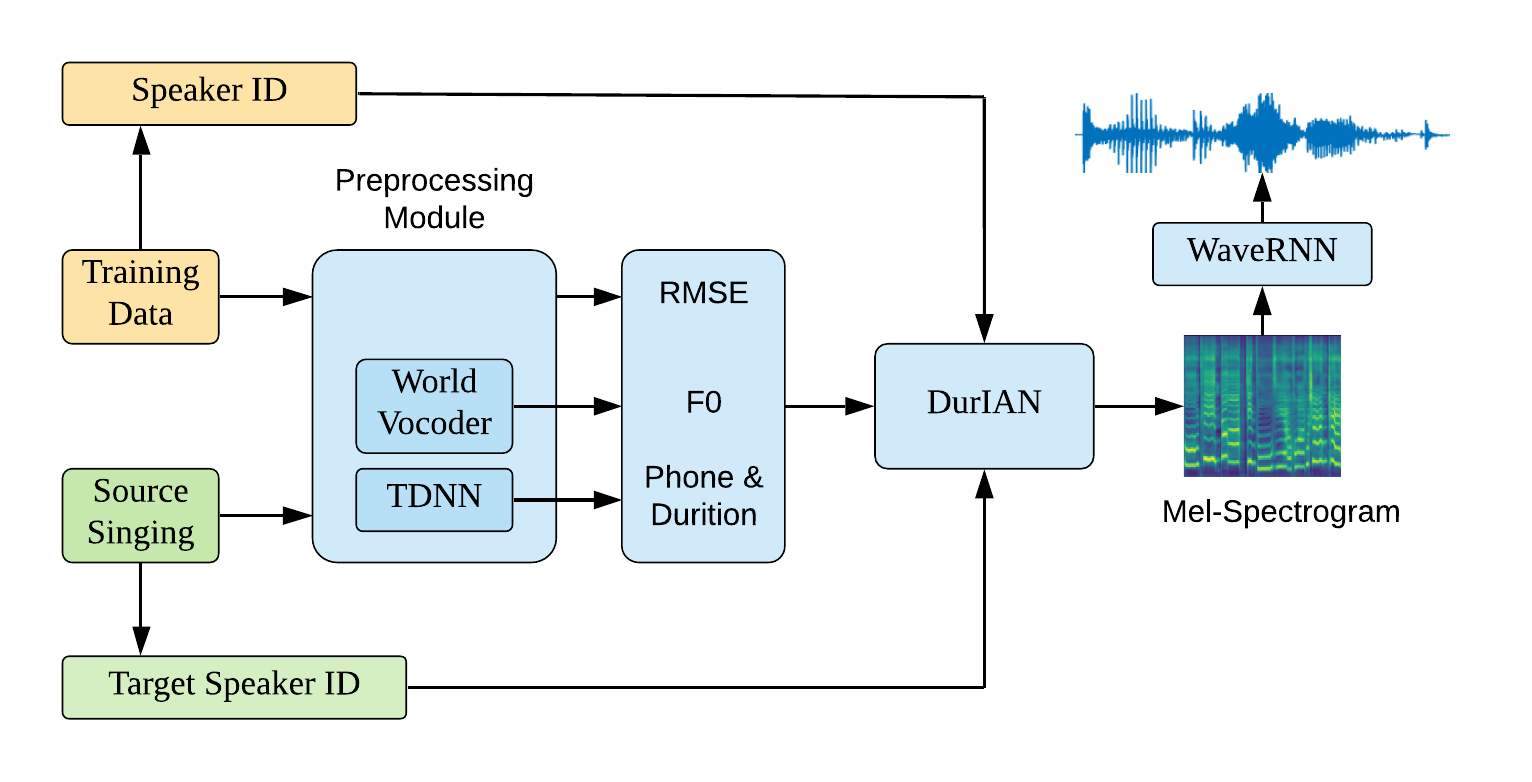}
\caption{The process diagram of training and converting. The yellow parts are used in training stage, the green parts are used in converting stage and the blue parts are used in both stages. The WaveRNN \cite{DBLP:journals/corr/abs-1802-08435} model is trained separately.}
\label{fig:f1}
\end{figure}

\subsubsection{Decoder} 
The decoder is the same as in DurIAN, composed of two autoregressive RNN layers. Different from the attention mechanism used in the end-to-end systems, the attention context is computed from a small number of encoded hidden states that are aligned with the target frames, which reduce artifacts observed in the end-to-end system. We decode two frames per time step in this paper. The output from the decoder network $\mathrm{y_{1:T}^{'}}$ is passed through a post-CBHG \cite{wang2017tacotron} to improve the quality of predicted mel-spectrogram as
\begin{equation}\label{eq6}
\mathrm{y_{1:T}^{'}=decoder(e_{1:T}^{'})}
\end{equation}
\begin{equation}\label{eq7}
\mathrm{\hat{y}_{1:T}=cbhg(y_{1:T}^{'})}
\end{equation}
The entire network is trained to minimize the mel-spectrogram prediction loss before and after post-CBHG as

\begin{equation}\label{eq8}
\mathrm{L=\sum_{i=1}^{T}|y_{i}-\hat{y}_{i}|+\sum_{i=1}^{T}|y_{i}-y_{i}^{'}|+l2loss}
\end{equation}
where $l2loss$ represents l2 regularization.

\subsection{Singing Voice Conversion}
\label{ssec:subhead}

The whole process of our method is illustrated in Fig. \ref{fig:f1}. The training dataset contains a multi-speaker speech and singing corpus. For singing voice conversion task, the target speaker or singer should be included in the training data, while the source singing or singer to be converted doesn’t have to be seen in the training. The preprocess module mainly consists of two parts: the TDNN based phoneme alignment model \cite{tdnn} and the world vocoder \cite{morise2016world}. The TDNN model is a component of a pre-trained general speech recognition model, which generates the phoneme sequence and its duration alignment from speech and singing data. The world vocoder is used to extract F0 which reflects the rhythm and melody of singing. Because the F0 envelope also determines the tone of each phone, so we use the  non-tonal phones in our experiment. In addition, we also found that the RMSE can greatly improve the quality and stability of singing voice conversion. The input of DurIAN-4S is phoneme sequence, phoneme durations, \textit{f0}, RMSE and speaker identity. The training target of DurIAN-4S is to reconstruct the mel-spectrogram. In the training stage, embeddings of speakers with speech samples and singing samples are all also optimized jointly.

After the model of DurIAN-4S is trained, it can be used to convert any singing to a target speaker's voice. The process of singing voice conversion is that, we first extract the f0, phoneme duration, and RMSE from the preprocess module, and use these as input for singing generation. By choosing different speaker embedding during singing generation, we could produce singing with different voice. The generated mel-spectrogram from DurIAN-4S after conversion will be used for WaveRNN \cite{DBLP:journals/corr/abs-1802-08435} model for waveform generation.   

When conversion between male and female, the input \textit{F0} should be multiplied by a scalar $\nu$ as:

\begin{equation}\label{eq1}
\nu  =  \frac{\sum_{i}^{N} \mathit{mean}(x_{i}^{t})}{N\cdot \mathit{mean}(x^{s})}
\end{equation}
where $x^{s}$ is the source singing, $x_{i}^{t}$ is the target speaker $\mathit{t}$, $\mathit{mean}$ is the average F0 of vowel phone in the audio. However, the pitch of one's singing is usually higher than the pitch of speech from the same person and it is common to adjust the key of songs within a certain range for different singers. We could control the scalar $\nu$ to get a flexible conversion performance.

\section{Experiment}
\label{sec:typestyle}

\subsection{Dataset}
\label{ssec:subhead}

The training set contains the Tencent multi-speaker speech corpus (TSP) and the Tencent singing corpus (TSG). In TSP corpus, we choose 3 male speakers and 4 female speakers, each with 1.5 hours of data. The TSG corpus contains a total of 28 hours singing data recorded by 3 female singers. For singing voice conversion task, we choose source singing from a separate singing corpus, which will not be used in training. All the data has a sampling rate of 24K.

\subsection{Model Parameters}
\label{ssec:subhead}

In our experiment, the dimensions of the phoneme embedding, speaker embedding, encoder CBHG module, attention layer are all 256. The decoder has 2 GRU layers with 256 dimension and the batch normalization is used in the encoder and post-net module. We use Adam optimizer and $0.001$ initial learning rate with warm-up \cite{goyal2017accurate} schedule. There is a total of 250,000 steps with a batch size of 32 to converge the model. We found multi-speaker trained WaveRNN model will improve the synthesis stability in this singing voice conversion task.

\subsection{Quality and Similarity Evaluation}
\label{ssec:subhead}
Since we are not able to find any public benchmarks on speech based singing voice conversion, we compared it singing voice conversion based on singing samples. Both the quality of converted singing voice and similarity of converted singing voice and target speaker's voice is compared. Subjective evaluation with Mean Opinion Scores (MOS) is used. A total of 14 subjects have been participated in our listening tests. 

We select 20 segments from two different songs from a separate singing corpus. Three male speakers and two female speakers from TSP corpus are selected as target speakers, and 1 singer from TSG corpus as target singer. We perform the experiments conduct an ablation study on the importance of using RMSE for singing voice conversion. For the timbre similarity evaluation, the subjects are asked to score a similarity of voice timbre between converted singing and target speaker's normal speech. 

The similarity evaluation results are shown in Table \ref{tab:mos}. The scale of MOS is set between 1 to 5 with 5 being the highest score. We first compare the effects of RMSE as additional input for singing voice conversion. And the results show that using RMSE improves both the quality and similarity significantly. 
We found that the energy information of each frame concatenated with F0 could help the model learning the pronunciation of long vowels. The energy of each frame indicates the loudness of pronunciation, helping the model to determine when vowels should stop properly.

We also compare the performance of singing voice conversion using target speaker's normal speech versus using their singing samples. Singing voice conversion using target speaker's singing samples receives better score than that of using their speech samples. This is expected performance, as it is much easier to learn speaker's singing voice from their singing samples than speech samples. However, we could see that the similarity of singing voice conversion using speech samples are not too far off, showing that proposed algorithm could synthesis target speaker's singing voice, in both high quality and similarity,  with only speech samples. The samples used in our experiments can be found at \url{https://tencent-ailab.github.io/learning_singing_from_speech}. 

\setlength{\abovecaptionskip}{-10pt}
\setlength{\belowcaptionskip}{-10pt}
\begin{table}[t]
\caption{MOS for Singing Conversion Quality and Similarity. Target speaker type indicates what types of samples we used for singing voice conversion from target speaker. 'singing' means the singing samples from target speaker are used for singing voice conversion, and 'speech' means speech samples from target speaker are used for singing voice conversion.}
	\begin{center}
		\label{tab:mos}
		{\small
		\begin{tabular}{cccc} 
			\toprule[1.5pt]
			\textbf{Method} & \textbf{Target Speaker} & \textbf{Naturalness} & \textbf{Similarity}\\
			\midrule
			f0 &singing & 3.23 & 3.00 \\
			\textit{f0} &speech & 2.77 & 2.84 \\
			\textit{f0} + RMSE &singing & 3.80 & 3.65 \\
			\textit{f0} + RMSE &speech & 3.42 & 3.49 \\
			\bottomrule[1.2pt]
		\end{tabular}
		}
	\end{center}
	\vspace{-4mm}
\end{table}

\section{Conclusion}
\label{sec:majhead}

In this paper, we proposed an algorithm that synthesizes natural singing in target speaker's voice given only their normal speech samples. We evaluate proposed algorithm on singing voice conversion task with speech samples, and obtained very promising results. In future work, we will focus on reducing the amount of target speech samples for both target singing synthesis and conversion tasks. 


\section{Acknowledgements}
\label{sec:majhead}

The authors would like to thanks Chunlei Zhang, Dongxiang Xu and other members in the Tencent AI Lab team for providing suggestions on model structure and optimization.

\vfill\pagebreak

\bibliographystyle{IEEEbib}
\bibliography{refs}

\end{document}